\documentstyle[a4,12pt,epsf]{article}
\def\be{\begin{equation}}
\def\ee{\end{equation}}
\def\bq{\begin{eqnarray}}
\def\eq{\end{eqnarray}}
\newcommand{\ra}{\rightarrow}

\begin{document}
\thispagestyle{empty}
\setcounter{page}{0}
\setcounter{page}{0}
\begin{flushright}
MPI-PhT/95-97\\
LMU 18/95\\
hep-ph/9510294 \\
October 1995
\end{flushright}
\vspace*{\fill}
\begin{center}
{\Large\bf B Meson Form Factors and Exclusive Decays}
$^{1,*}$\\
\vspace{2em}
\large
A. Khodjamirian$^{a,\dagger}$ and R. R\"uckl$^{a,b}$\\
\vspace{2em}
{$^a$ \small Sektion Physik der Universit\"at
M\"unchen, D-80333 M\"unchen, Germany }\\
{$^b$ \small Max-Planck-Institut f\"ur Physik, Werner-Heisenberg-Institut,
D-80805 M\"unchen, Germany}\\

\end{center}
\vspace*{\fill}

\begin{abstract}
We present results for the
$B \rightarrow \pi$ and $D\rightarrow \pi$ form factors derived from
QCD sum rules on the light-cone. Our predictions are compared
with experiment and used to extract the quark mixing
parameter  $|V_{ub}|$ from a recent CLEO measurement of
$B \ra \pi l^+ \nu$.
Furthermore, we discuss the factorization approximation for
exclusive nonleptonic matrix elements, and describe
a first QCD estimate of the nonfactorizable
contribution to the amplitude of $B \rightarrow J/\psi K $.

\end{abstract}

\vspace*{\fill}

\begin{flushleft}

\noindent$^\dagger$on leave from Yerevan Physics Institute,
375036 Yerevan, Armenia \\
\noindent$^1${\small supported by the German Federal Ministry
BMBF under contract 05 6MU93P,
and by the EC-program HCM under contract CHRX-CT93-0132}
\\
\noindent$^*${\it invited talk presented by R. R\"uckl at the
Workshop `Beauty 95', Oxford,
July 9-14,1995, to appear in the proceedings}
\baselineskip=16pt
\end{flushleft}

\newpage
\section{Introduction}
Heavy flavour decays play an outstanding role in the determination
of fundamental parameters of the standard model and for the
development of a deeper understanding of the QCD dynamics.
In order to achieve the ambitious scientific goals
it is important to combine the experimental
and theoretical efforts. However, this is not always easy.
For example, while the theory of inclusive decays is most
advanced, inclusive measurements are generally quite difficult.
Conversely, exclusive decays into few-body final
states are often much easier to detect and reconstruct,
but the theoretical methods to treat exclusive processes are still
rudimentary when hadrons are involved. In view of the exciting
experimental prospects at future bottom and charm factories,
where many new
exclusive channels will become accessible, it
appears particularly desirable to have a
reliable quantitative theory.

As well known, the main problem is the complicated interplay
of weak and strong interactions,
and most of all the influence of long-distance dynamics.
Clearly, a complete theoretical understanding of the
decays of hadrons containing a heavy quark is only
possible in a nonperturbative framework.
This is
most obvious for exclusive nonleptonic decays and for
processes involving strong
couplings among hadrons such as $B^*B\pi$ or $D^*D\pi$.
Yet, what appears as a problem for determining electroweak
parameters from exclusive measurements can also be regarded as a
chance to improve and test our understanding of
hadronic physics in QCD.

The nonperturbative approaches applied in this field include
lattice approximations, heavy quark effective theory,
chiral perturbation theory,
$1/N_c$-expansion, QCD sum rules, and quark models.
Each of these methods has advantages and disadvantages.
For example, quark models are easy to use and very good
for intuition, however, their relation to QCD is unclear.
In the other extreme, lattice calculations are rigorous from the
point of view of QCD, but they can (at least presently)
be applied only to relatively simple matrix elements related to decay
constants and form factors. Moreover, there are uncertainties
connected with the necessary extrapolations to the physical
masses of the light (u,d) and heavy (b) quarks.
Also the application of effective theories is restricted
to a rather narrow class of problems, and the corrections
are often substantial. More importantly, in many cases
the latter  cannot be calculated directly within the same theory,
and therefore bring in model-dependence after all.

QCD sum rules \cite{SVZ}
represent no exception is this respect.
On the one hand, they are based on first principles.
On the other hand,
one has to introduce new elements such as vacuum condensates
or light-cone wave functions which cannot (yet) be
calculated directly in QCD. However, since
this nonperturbative input is of universal nature,
sum rules nevertheless possess a high predictive power.
Furthermore, the sum rule approach is particularly
well suited for heavy quark physics, since
the approximations which have to be made in practice
can be well justified when a heavy mass scale is present.
Last but not least, the sum rule approach is less
limited in applicability
than some of the other methods mentioned, or
the limitations are different.
In this paper we will exemplify some recent applications of the
QCD sum rule technique to $B$ and, for comparison, also to
$D$ decays.

The simplest and therefore most instructive example for
introduction is actually provided by the calculation of
heavy meson decay constants. They are defined by
matrix elements of quark currents, for example
\be
m_b\langle B \mid\bar{b}i\gamma_5 d\mid 0\rangle =f_Bm_B^2~,
\label{fB}
\ee
and determine leptonic decays such as $B \ra \tau \nu_{\tau}$.
In addition they enter in the factorizable amplitudes of
nonleptonic decays. Without going into details, which for example
can be found in \cite{RRY}, we only mention that the QCD sum
rule approach to (\ref{fB}) is based on an analysis of the
two-point correlation function
$\langle 0\mid T\{\bar{b}(x)i\gamma_5 d(x),
\bar{d}(0)i\gamma_5 b(0)\}\mid 0\rangle$ using
operator product expansion, perturbation theory, and
dispersion relations. $T$ in the above current-correlator
denotes the time-ordering operation.

For brevity, we directly proceed to more complicated
cases and begin with the calculation of form factors. In section 2,
we describe how sum rules on the light-cone can be used to
derive the $B \ra \pi$ and $D \ra \pi$ form factors. Comparison
with recent experimental results on
CKM-suppressed $B$ and $D$ decays as well as the determination of
$|V_{ub}|$ are discussed in section 3.
Finally, section 4 is devoted to the main theoretical problem
in heavy quark decays that is the complete calculation of
exclusive nonleptonic amplitudes. We briefly demonstrate the failure
of strict factorization, the usual phenomenological
procedure, and a first attempt to estimate the nonfactorizable
piece of the amplitude for $B\ra J/\psi K$.

\section{The $B \ra \pi$ and $D \ra \pi$ form factors }

Recently, the CLEO collaboration \cite{CLEOBpi}
announced the first measurement of the decay   $B\rightarrow \pi l\nu$.
This process plays a very important role
for the determination of the
CKM parameter $V_{ub}$. The decay amplitude is completely
determined by the form factor
$f^{+}_{B}(p^2)$ entering the matrix element
\be
<\pi(q)|\bar{u} \gamma_\mu b |B(p+q)> =
2f^{+}_{B}(p^2)q_\mu +\left[f^{+}_{B}(p^2) +
f^{-}_{B}(p^2)\right]p_\mu,
\label{matrix}
\ee
where the assignment of the momenta $p$ and $q$ is quite obvious.
In \cite{BKR} this form factor
was calculated from a QCD sum rule on the light-cone.
In the following we briefly outline the method.

One starts from the correlation function of the
relevant vector and pseudoscalar currents
between the vacuum and the on-shell pion state:

\bq
F_\mu(p,q)= i\int d^4x e^{ipx}
\langle\pi(q)\mid T\{\bar{u}(x)\gamma_\mu b(x),
\bar{b}(0)i\gamma_5 d(0)\}\mid 0\rangle
\nonumber
\\
= F(p^2,(p+q)^2)q_\mu + \tilde{F}(p^2,(p+q)^2)p_\mu  ~.
\label{1}
\eq
Writing a dispersion relation in the variable
$(p+q)^2$, one readily finds the $B$ meson ground state
contribution to the invariant function $F(p^2,(p+q)^2)$
by inserting a complete sum of states between the currents
in (\ref{1}), focusing on the term
$ |B \rangle \langle B | $
and using the relations (\ref{fB}) and (\ref{matrix}):
\be
F(p^2,(p+q)^2)= \frac{2m_B^2f_Bf_B^+(p^2)}{m_b(m_B^2-(p+q)^2)}+\ldots~.
\label{disp}
\ee
The ellipses in the above denote
the contributions from the excited $B$ and from continuum states.

The main idea of the method is to
apply the operator product expansion (OPE)
to the product of currents in the correlation function (\ref{1})
at spacelike momentum $(p+q)^2<0$
such that  the b-quark propagating from point $x$ to point 0
is highly virtual. Moreover, the expansion is performed near
the light-cone, $x^2=0$ , rather than around $x=0$. Consequently,
one ends up with a series of bilocal rather than local
operators. More definitely,
after contracting the $b$-quark fields in (\ref{1})
and expanding the $b$-quark propagator near $x^2=0$,
the correlation function  can be expressed in terms of
quark-antiquark and quark-antiquark-gluon
matrix elements of the type
\be
\langle \pi(q) \mid \bar{u}(x) \Gamma_a d(0) \mid 0 \rangle~~
\mbox{and} ~~\langle \pi(q)
\mid \bar{u}(x) g_sG^{\mu\nu}(vx)\Gamma_b d(0) \mid 0 \rangle~,
\label{matrixel}
\ee
respectively, where $ \Gamma_{a,b}$ denote certain combinations
of Dirac matrices and the auxiliary variable $v$ varies from 0 to 1.
These matrix elements are essentially nonperturbative objects.
Most importantly, they are universal and  process-independent.
In particular, the same matrix elements enter in the
corresponding expansion of the correlation function
interpolating the $D\ra \pi$ form factor. The latter
can be immediately obtained from (\ref{1})
by replacing $b$ by $c$ and $B$ by $D$.

The vacuum-to-pion matrix elements introduced above
can be parameterized
in terms of so-called light-cone wave functions
with given twist \cite{BL,CZ,BF}.
The most important of them is the leading
twist 2 wave function $\varphi_\pi(u,\mu) $
defined by
\be
\langle\pi(q)|\bar{u}(x)\gamma_\mu\gamma_5d(0)|0\rangle=
-iq_\mu f_\pi\int_0^1du\,e^{iuqx}\varphi_\pi (u,\mu)+ ...  ~,
\label{pionwf}
\ee
where $\mu ^{-1} $ is a characteristic distance scale
and $u$ is the fraction of the pion light-cone momentum
$q_0 +q_3$  carried by the constituent quark.
The ellipses in (\ref{pionwf}) stand for the higher twist components.
Asymptotically, that is at $\mu\ra \infty$, QCD
perturbation theory implies
$ \varphi_\pi(u,\infty) = 6u(1-u)$. However, at the physical scale
$\mu \sim m_b $, at which the OPE is applied to the
correlation function (\ref{1}),
important nonasymptotic effects are to be expected.
These are parameterized by the coefficients
$a_i(\mu)$ in the following series of Gegenbauer
polynomials:
\be
\varphi_\pi(u,\mu)=6u(1-u)[1+\sum_{i=2,4,..} a_i(\mu)C^{3/2}_i(2u-1)] ~.
\label{u}
\ee
Although the input values
of the coefficients $a_i(\mu)$ at a fixed low momentum scale $\mu$
are unknown,
the scale-dependence of $a_i(\mu)$ is dictated by the renormalization
group. From what is said above, it is also clear
that $a_i(\mu) \ra 0$ for  $ \mu \ra\infty$.
Over the years a great deal has been learned about these wave
functions. They have been classified
up to twist 4 and the asymptotic form has been determined.
Also various nonasymptotic corrections have been estimated from
QCD sum rules for light hadrons. In our calculation
we have used the wave functions given in \cite{BF}.

Substitution of (\ref{pionwf}) into the light-cone expansion of the
correlation function (\ref{1})  yields a
representation of the invariant amplitude $F$ in terms
of perturbative and nonperturbative parameters and functions
of QCD. The leading term of it reads
\be
F( p^2,(p+q)^2)=
m_bf_\pi\int_0^1\frac{du ~\varphi_\pi(u,\mu) }{m_b^2-(p+uq)^2}~
+ \ldots ~.
\label{qcd}
\ee
Equating now the QCD result (\ref{qcd}) in the region
of validity at $(p+q)^2<0$ with
the dispersion relation (\ref{disp}) one obtains
a raw sum rule relation for $f_Bf^+_B(p^2)$.
The rest of the calculation follows the usual QCD sum rule
procedure: Borel transformation  in $(p+q)^2$ and subtraction
of the contribution from higher states invoking
quark-hadron duality. Details can be found in the original
paper \cite{BKR}.
One finally arrives at an expression for the
desired form factor, the dominant term of which is given by
\be
f^+_B( p^2)= \frac{f_\pi m_b^2}{2f_Bm_B^2}\int_
\Delta^1\frac{du}{u}(\varphi_\pi(u,\mu_b) + \ldots)
exp \left(\frac{m_B^2}{M^2}
-\frac{m_b^2-p^2(1-u)}{uM^2}\right) ~.
\label{formf}
\ee
Here, $M$ is the Borel mass parameter, and
the scale $\mu_b$ is of order of the characteristic
virtuality of the correlation function, $\mu_b^2 = m_B^2-m_b^2$.
The integration limit $\Delta = (m_b^2-p^2)/(s_0-p^2) $
depends on the effective threshold $s_0$ above which the
contribution from higher states to the dispersion relation
(\ref{disp}) are cancelled against the corresponding piece
in the QCD representation (\ref{qcd}).
The twist 3 and 4 contributions denoted by ellipses have
also been calculated in
\cite{BKR,BBKR}, but are not shown here for brevity.
They involve further quark-antiquark wave
functions as well as three-particle
quark-antiquark-gluon wave functions.

An important detail which should be mentioned here concerns
the treatment of $f_B$ in the sum rule (\ref{formf}).
For consistency, $f_B$
is replaced by the appropriate two-point sum rule,
disregarding the $O(\alpha_s)$ gluon
corrections since the latter are also not yet included in
(\ref{formf}). This procedure considerably decreases
the sensitivity of $f_B^+$ to the choice of $m_b$ and $s_0$
as demonstrated in \cite{BKR,BBKR}.
Furthermore, we observe a remarkable stability of (\ref{formf})
with respect to a variation of the
Borel parameter $M$. The appropriate range
of $M^2$ is found by requiring the higher twist
terms and simultaneously,
the contributions from heavier states to remain subleading.

In Fig. 1a the form factor $f^+_B(p^2)$ is plotted  versus $p^2$
for the central value  $M^2=10$ GeV$^2$ of
the fiducial range.
On general grounds, one can expect the sum rule (\ref{formf})
to be valid up to a timelike momentum transfer
$p^2 = m_b^2-O$(1 GeV$^2$). Actually, the calculation shows that above
$p^2 \simeq 17\div 20$ GeV$^2$ the stability in $M^2$ is lost.
In Fig. 2 we compare our prediction with the results
obtained in a quark model \cite{BSW} and from a three-point
sum rule involving local quark and
gluon condensates instead of light-cone wave functions \cite{BBD}.

Obvious replacements in (\ref{1}) and (\ref{formf})
yield the corresponding sum rule
for the $D \ra \pi$ form factor. The
numerical result for $f^+_D(p^2)$ is plotted in Fig. 1b
taking $M^2= 4 $ GeV$^2$ in accordance with the reliability criteria
pointed out above. Note again that this calculation should
only be trusted at $p^2 <m_c^2 - O(1 \mbox{GeV}^2) $.

For easy comparison with  other theoretical results
(a rather complete compilation can be found in \cite{Ball} )
we also quote our predictions at $p^2=0$:
\be
f^+_B(0)= 0.29\pm 0.01~~,~~
f^+_D(0)= 0.66 \pm 0.03~~.
\label{zero}
\ee
The uncertainty above only refers to the variation
with $M^2$ within the fiducial range.
We have also investigated the sensitivity
to nonasymptotic effects in the light-cone wave functions.
To this end we have recalculated $f^+_B$ and $f^+_D$
using purely asymptotic wave functions and find that the
form factors change by less than
10 \% . This gives additional confidence in the results exhibited
in Fig. 1.

For an accurate theoretical analysis of the semileptonic
decays $ B \ra \pi l \nu $ and $ D \ra \pi l \nu $
one also has to know the form factors at larger
values of $p^2$, ideally  up to the zero recoil points
$( m_B-m_\pi)^2$ and $( m_D-m_\pi)^2$, respectively.
At such large $p^2$
the form factors are expected to be dominated by the $B^*$ and
$D^*$ poles, respectively. In this approximation,
\be
f^+_B(p^2)= \frac{f_{B^*}g_{B^*B\pi}}{2m_{B^*}(1-p^2/m_{B^*}^2)}~,
\label{onepole}
\ee
where $g_{B^*B\pi}$ is the $B^*B\pi$ coupling defined by
\be
\langle B^{*-}(p)\pi^+(q)\mid \bar{B}^0(p+q)\rangle =
-g_{B^*B\pi}q_\mu \epsilon ^{\mu}~.
\label{BBpi}
\ee

It is one of the advantages of the light-cone sum rule method
that the $B^*B\pi$ coupling can be calculated \cite{BBKR} from
the same correlation function (\ref{1}).
One only has to set up sum rules with the respect to
both the $B$- and $B^*$-
channel considering a dispersion relation in
$(p+q)^2$ and $p^2$, simultaneously.
After double Borel transformation of the invariant amplitude
(\ref{qcd}) involving the mass parameters $M_1$ and $M_2$
one obtains an expression for
the product $f_Bf_{B^*}g_{B^*B\pi}$, the leading twist term of which
depends on the pion wave function
$ \varphi_\pi(u,\mu)$ at the fixed momentum fraction
$ u = M_1^2/(M_1^2+M_2^2) \simeq 1/2 $:
\be
f_Bf_{B^*}g_{B^*B\pi}=\frac{m_b^2f_\pi}{m_B^2m_{B^*}}
e^{\frac{m_{B}^2+m_{B*}^2}{2M^2}}
\Bigg\{M^2 [e^{-\frac{m_b^2}{M^2}} - e^{-\frac{s_0}{M^2}}]
\varphi_\pi(1/2)+ \ldots \Bigg\}
\label{g}
\ee
with $M^2= M_1^2M_2^2/(M_1^2+M_2^2)$.
Taking $ \varphi_\pi(1/2, \mu=\mbox{0.5 GeV})=1.2 \pm 0.2 $ \cite{BF}
together with the corresponding values of the
higher twist wave functions and dividing by the values of
$f_B$ and $f_{B^*}$ as given by the appropriate two-point sum rules,
we find, numerically,
\be
g_{B^*B\pi}= 29\pm 3 ~,
\label{coupl}
\ee
where the indicated error again quantifies the variation within
the fiducial range of the Borel parameters.
%%%
% ffffffff---Fig.1---fffffffffffffffffffff
%\newpage
\vspace*{4.8cm}
\begin{center}
\begin{minipage}[t]{7.8cm} {
%----- Fig. 1 ---
\begin{center}
%----------------
\hspace{-3.2cm}
\mbox{
\epsfysize=10cm
\epsffile[0 0 500 500]{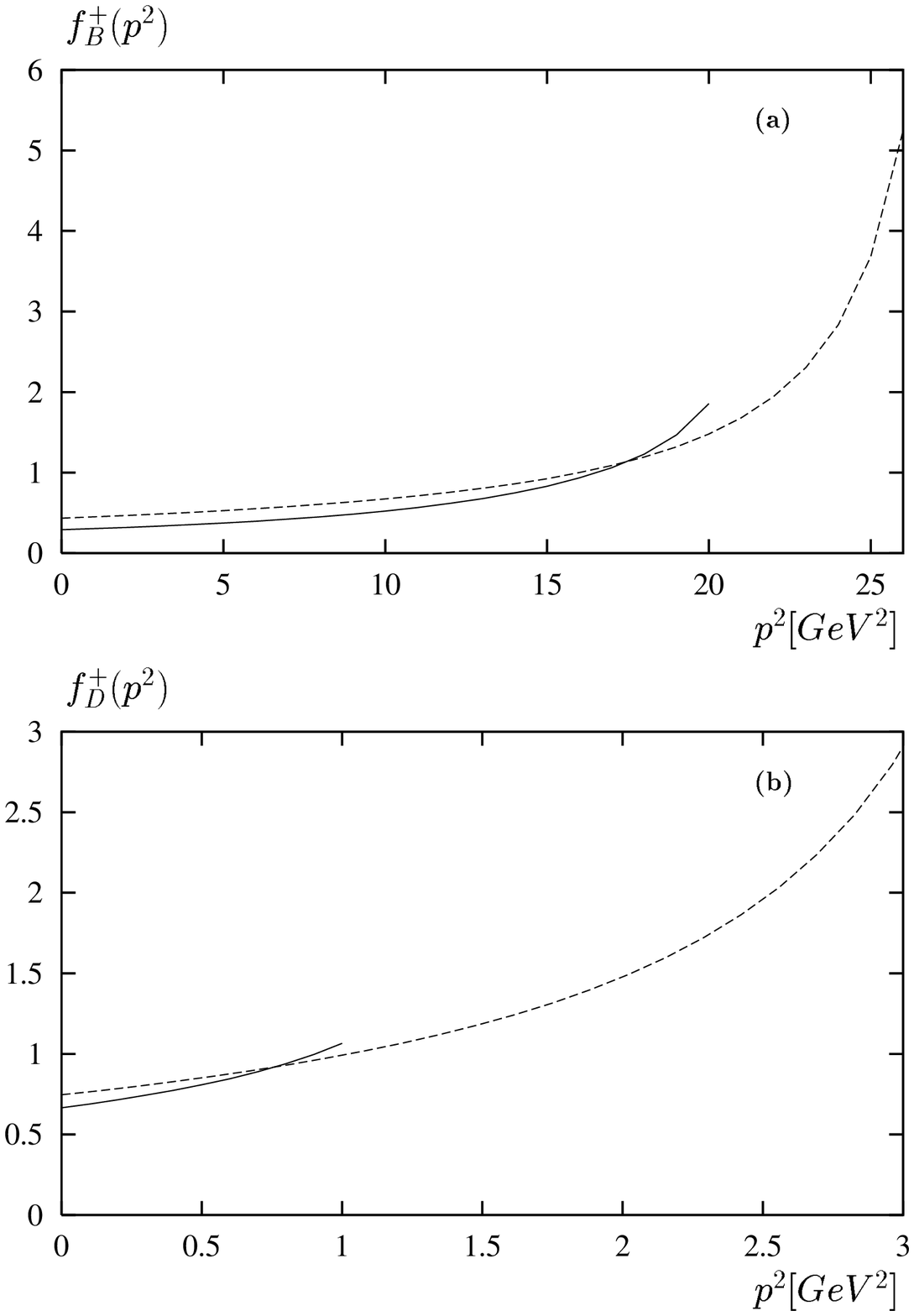}
}
%----------------
\end{center}
%\vspace*{-2.0cm}
%\noindent
}\end{minipage}
\end{center}
%\vspace{0.5cm}
\vspace{-3.0cm}
%%%%%%%%%%%%%%%%%%%%
{\small\bf Fig.~1. }{\small\it The form factors for the transitions
(a) $B \ra \pi$  and (b) $D \ra \pi$
as predicted by light-cone sum rules (solid)
in comparison to the single-pole approximation
(dashed) with the normalization
constants $g_{B^*B\pi}$ and $g_{D^*D\pi}$, respectively,
calculated by the same method.}
\bigskip

Expressions analogous to (\ref{onepole}) and  (\ref{BBpi})
hold for  $f^+_D(p^2)$ in the region near
$p^2 \simeq (m_D-m_\pi)^2 $. Following the same procedure as the one
described above, we get
\be
g_{D^*D\pi}= 12.5\pm 1.0~.
\label{couplD}
\ee
This prediction implies the partial width
\be
\Gamma (D^{*+} \ra D^0 \pi^+ )=32 \pm 5~ \mbox{keV}
\label{Dwidth}
\ee
% ffffffff---Fig.2---fffffffffffffffffffff
%\newpage
\vspace*{1.5cm}
\begin{center}
\begin{minipage}[t]{7.8cm} {
%----- Fig. 1 ---
\begin{center}
%----------------
\hspace{-3.2cm}
\mbox{
\epsfysize=10cm
\epsffile[0 0 500 500]{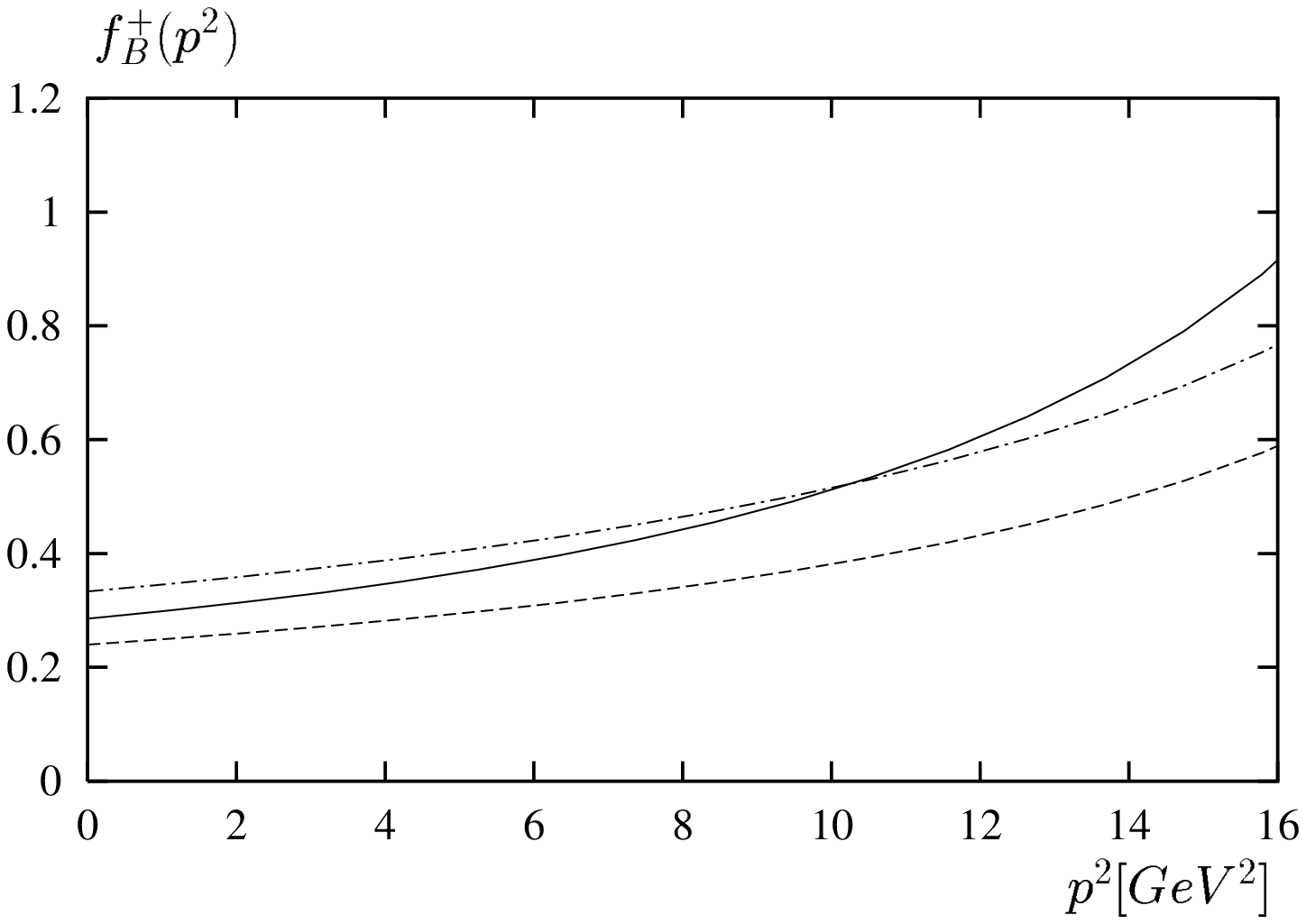}
}
%----------------
\end{center}
}\end{minipage}
\end{center}
%\vspace{0.5cm}
\vspace{-5.0cm}
%%%%%%%%%%%%%%%%%%%%
{\small\bf Fig.~2. }{\small\it
Comparison of the $B\ra \pi$ form factor
as derived from the light-cone sum rule (solid)
with the predictions of a three-point sum rule
\cite{BBD} (dashed) and a quark model \cite{BSW}
(dash-dotted).  }
\bigskip

\noindent which is consistent with the limit derived from recent
ACCMOR \cite{ACCMOR} and CLEO \cite{CLEO1} measurements.

The single-pole approximation for the form factors $ f^+_B(p^2)$
and  $f^+_D(p^2)$  obtained with (\ref{coupl}) and (\ref{couplD})
is shown in Fig. 1a and Fig. 1b, respectively.
Whereas the direct estimate of the form factors such as
(\ref{formf})
is trustful at low and intermediate values of $p^2$,
but breaks down as $p^2 \ra m^2_b$ and $m_c^2$, respectively,
there are no convincing arguments in favour of
the pole model to be valid at $p^2 \ra 0$.
Hence we suggest to match
the two descriptions in the region of intermediate momentum transfer
near the upper end of the full curves plotted in Fig. 1. More precisely,
in the following applications we use the direct
calculation of $f^+_B$ ($f^+_D$)
up to $p^2=15~(1.0) $ GeV$^2$
and the single-pole approximation
at higher values of $p^2$.

In order to improve these calculations further one has to
take into account the $O(\alpha_s)$
hard gluon exchanges in the correlation
function (\ref{1}) explicitly. Parts of this not quite
straightforward
task have already been completed \cite{Weinzierl}.

\section{Extraction of $V_{ub}$ from $B \ra \pi l \nu$ }
With the form factor $f^+_B(p^2)$ at hand, we are now
in the position to determine the CKM parameter  $|V_{ub}|$ from
the recent CLEO measurement of $B^0 \ra \pi^- l^+ \nu_l$.
Fig. 3a shows the charged lepton energy spectrum
in the $B^0$ rest frame,
\bq
\frac{d\Gamma(B^0\rightarrow\pi^- l^+ \nu_l)}{dE_l}=
\frac{G^2|V_{ub}|^2}{16\pi^3m_B}\int^{p_{max}^2}_0 dp^2
\Bigg[ 2E_l(m_B^2-m_\pi^2+p^2)
\nonumber
\\
-m_B(p^2+4E_l^2)\Bigg]
\left[f^+_B(p^2)\right]^2  ~,
\label{spectrum}
\eq
with $p^2_{max}(E_l) = 2E_l(m_B-m_\pi^2/(m_B-2E_l))$.
Integration of the spectrum over $E_l$ yields  the
exclusive decay width
\be
\Gamma(B^0\rightarrow\pi^- l^+ \nu_l) =
8.1 ~|V_{ub}|^2~\mbox{ps}^{-1}~.
\label{Bwidth}
\ee

Experimentally, combining the preliminary
CLEO result,
$BR(B^0\ra\pi^-l^+\nu_l) = (1.63 \pm 0.46\pm 0.34)\cdot 10 ^{-4}$~
\cite{CLEOBpi},
with the world average of the $B^0$ lifetime
\cite{Rizzo},  $\tau_{B^0}=1.57 \pm 0.05 $ ps,
one obtains
\be
\Gamma(B^0\rightarrow \pi ^- l^+ \nu_l) =
(1.04 \pm 0.37)\cdot10^{-4}~\mbox{ps}^{-1}~,
\label{Gammaexp}
\ee
where the errors have been added in quadrature.
We take the CLEO result obtained by using the BSW model \cite{BSW}
for the efficiency estimate since the BSW  form factors
agree quite well with our predictions.

Comparison of (\ref{Bwidth}) with (\ref{Gammaexp})
yields
\be
|V_{ub}| = 0.0036 \pm 0.0006~.
\label{Vub}
\ee
The theoretical uncertainty which we estimate
conservatively to be less than
20\%  is not included in (\ref{Vub}).

% ffffffff---Fig.3---fffffffffffffffffffff
%\newpage
\vspace*{4.8cm}
\begin{center}
\begin{minipage}[t]{7.8cm} {
%-----
\begin{center}
%----------------
\hspace{-3.2cm}
\mbox{
\epsfysize=10cm
\epsffile[0 0 500 500]{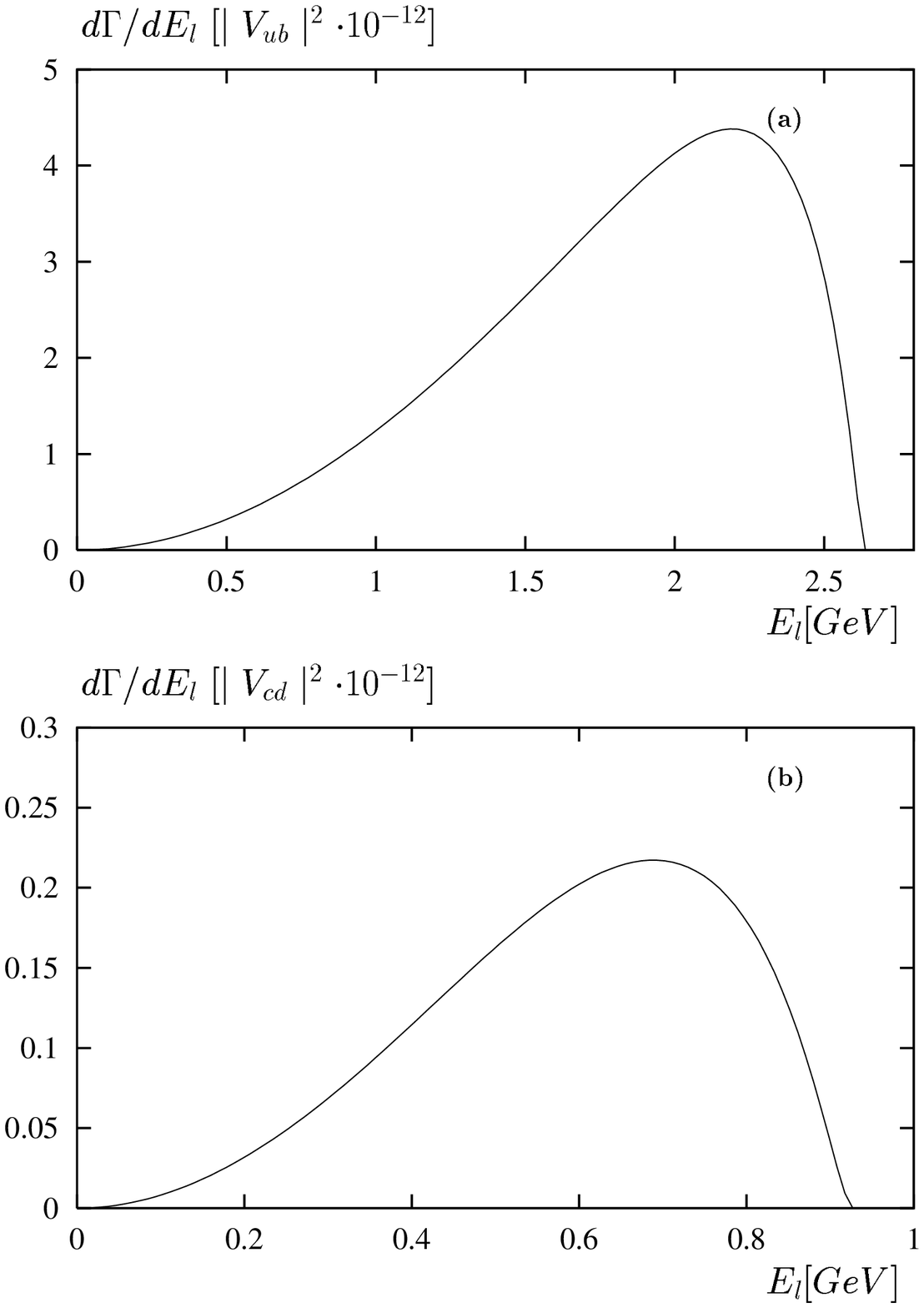}
}
%----------------
\end{center}
%\vspace*{-2.0cm}
%\noindent
%{\small\bf Fig.~1. }{\small\it
%
%}
}\end{minipage}
\end{center}
%\vspace{0.5cm}
\vspace{-3.0cm}
%%%%%%%%%%%%%%%%%%%%
{\small\bf Fig.~3. }{\small\it
The charged
lepton spectra in (a) $B^0 \ra\pi^- l^+ \nu $ and
(b) $D^0 \ra\pi^- l^+ \nu $ .}
\bigskip

In the case of the Cabibbo suppressed $D$ meson decay
$D^0\rightarrow \pi^- l^+ \nu_l $
we predict the charged lepton energy spectrum shown in Fig. 3b
and the integrated semileptonic width
\be
\Gamma(D^0\rightarrow\pi^- l^+ \nu_l) =
0.156 ~|V_{cd}|^2 ~\mbox{ps}^{-1} =
(7.6 \pm 0.2)\cdot 10^{-3} ~\mbox{ps}^{-1}~,
\label{Gamma}
\ee
where we have substituted
$|V_{cd}|= 0.221 \pm 0.003 $ \cite{PDG}.

This prediction should be compared with
the experimental result
\be
\Gamma(D^0\rightarrow\pi^- e^+ \nu_e) =
(9.4^{+5.5}_{-2.9})\cdot10^{-3} ~\mbox{ps}^{-1},
\label{brD}
\ee
derived from
the branching ratio
$BR(D^0\rightarrow\pi^- e^+ \nu) =
(3.9^{+2.3}_{-1.2})\cdot 10^{-3}$
and the lifetime
$\tau_{D^0}= 0.415 \pm 0.004  ~\mbox{ps}$ ~\cite{PDG}.
The present theoretical uncertainty in
(\ref{Gamma}) is of the order of the experimental error in
(\ref{brD}).

We see that  the
CKM-suppressed exclusive semileptonic
widths are not  yet  measured
precisely enough to really challenge theory.
In our approach,
when better data become available, one may
reduce the remaining uncertainties
considerably and determine the CKM matrix elements with
better accuracy.
The idea is simple. Both the $B$ and $D$ meson
sum rules are determined
by one and the same set of nonperturbative parameters,
including the coefficients of the
nonasymptotic terms in the light-cone wave functions
as exemplified in (\ref{u}).
The present uncertainty in these parameters may be further
reduced by constraining them through a
precise measurement of
$ D\ra \pi l\nu$. After evolution from $\mu \sim m_c$
to $\mu \sim m_b$, one should then obtain a more
accurate prediction on $B\ra \pi l \nu$ and thus improve
also the result on $|V_{ub}|$.

In conclusion we mention that
the second form factor $f^-_\pi(p^2)$ of the  $B\ra \pi$
transition matrix element (\ref{matrix}) can be calculated
by the same method. The work is in progress \cite{KRW}.
This form factor may become measurable in future
experiments in the decay $B\ra \pi \tau\nu_\tau $.

\section{ The nonfactorizable $B \ra J/\psi K $ amplitude}

To proceed from semileptonic to nonleptonic decays
when QCD is switched off, one simply has to replace the
lepton current by the
relevant quark current in the weak Hamiltonian and multiply
the latter  by the corresponding CKM matrix element. Of course, the
strong interaction effects
then become much more complicated. They arise from
(a) hard gluon exchange at short distances,
(b) soft interactions of quarks and gluons including
nonspectator effects, (c)
the confinement mechanism leading
to the formation of hadrons, and
(d) final state interactions among the hadronic
decay products.

Theoretically, only the effect (a) can be systematically taken
into account in the framework of improved QCD perturbation
theory. The result is an
effective weak Hamiltonian at the physical scale $\mu << m_W$
of interest, given by a sum of local operators
with renormalized Wilson coefficients \cite{Buras}.
In the following we
consider the decay $B \rightarrow J/\psi K $ as an
important and instructive study case. The part of the
effective Hamiltonian relevant for this decay
may be written in the form
\be
H_W= \frac{G}{\sqrt{2}}V_{cb}V^*_{cs}\{(c_2
+\frac{c_1}3) O_2+2c_1\tilde{O}_2\}~,
\label{H}
\ee
where
\be
O_2(\mu)=(\bar{c}\Gamma^\rho c)(\bar{s}\Gamma_\rho b),\
\tilde{O}_2(\mu)=(\bar{c}\Gamma^\rho \frac{\lambda^a}2c)(\bar{s}\Gamma_\rho
\frac{\lambda^a}2 b)
\label{o}
\ee
with $\Gamma_\rho = \gamma_\rho(1-\gamma_5)$ and $\lambda ^a$ being the
usual SU(3) colour matrices.
The Wilson coefficients $c_{i}(\mu)$ contain the QCD corrections from
short distances below a characteristic distance scale
of the order of the inverse $b$-quark mass.
The remaining task is then to calculate
the hadronic  matrix elements of the four-quark operators (\ref{o})
incorporating the effects (b) to (d). This clearly requires
nonperturbative methods whence progress has been slow and difficult.

In the most radical approach,
the matrix elements of $H_W$ are factorized into products of
hadronic matrix elements of the currents that compose $H_W$.
Strong interactions at scales lower than $\mu$
between quarks entering different
currents as well as nonspectator effects are completely neglected
in this approximation. Hence, the resemblance of
nonleptonic decays to
semileptonic decays goes quite far. Formally,
the matrix element of the operator
$\tilde{O}_2$ vanishes  because of colour conservation,
and the amplitude for $B(p+q) \rightarrow J/\psi(p)~ K(q) $
(momenta in parenthesis) is given by
the factorized matrix element of the operator $O_2$:
\be
\langle J/\psi K\mid H_W\mid B\rangle
= \frac{G}{\sqrt{2}}V_{cb}V^*_{cs} \left(c_2(\mu)
+\frac{c_1(\mu)}3 \right)
\langle J/\psi K \mid O_2(\mu) \mid B\rangle~
\label{factO}
\ee
with
$$
\langle J/\psi K\mid O_2(\mu)\mid B\rangle
= \langle J/\psi\mid \bar{c}\Gamma^\rho c  \mid 0 \rangle
\langle K  \mid \bar{s}\Gamma_\rho b \mid B\rangle
$$
\be
= 2f_\psi f_{B \ra K}^+m_\psi(\epsilon^\psi  \cdot q) ~.
\label{factoriz}
\ee
While the decay constant $f_{\psi}$ can be
obtained from the leptonic width $\Gamma( J/\psi$\-$ \ra l^+l^-)$,
the $B\ra K$ form factor $f_{B \ra K}^+( p^2=m_\psi^2)$
can be calculated in analogy to the $B \ra \pi$
form factor discussed in Section 2 .
The result used below is $f_{B \ra K}^+(m_\psi^2) = 0.55 \pm 0.05 $
\cite{BKR}. Obviously, $\epsilon^\psi$ denotes
the $J/\psi$ polarization vector.
At this point one encounters a principal problem:
since the matrix elements  of quark currents in (\ref{factoriz})
are scale independent, the $\mu$-dependence  of
$\langle J/\psi K \mid O_2(\mu) \mid B\rangle $ which may
cancel the $\mu$-dependence of the Wilson coefficients
in (\ref{factO}) and give a
physically sensible result, is lost (see also the discussion
in \cite{Buras}).
Hence the approximation (\ref{factoriz}) can at best be
valid at a particular value of $\mu$ which could be called
the factorization scale $\mu_F$. Usually, $\mu_F$ is expected
to be of the order of the heavy quark mass.

Using the next-to-leading coefficients $c_{1,2}(\mu)$
in the HV scheme for
$\Lambda^{(5)}_{\overline{MS}}=225$ MeV from
\cite{Buras2} and  taking $\mu=m_b \simeq 5 $ GeV,
one obtains $a_2= 0.155$ and from that the branching ratio
\be
BR( B\rightarrow J/\psi K) = 0.025\% .
\label{BRfact}
\ee
This estimate is significantly
smaller than the measured branching ratios
\cite{PDG,CLEO}:
\be
BR( B^- \rightarrow J/\psi K^- )= (0.102 \pm 0.014 )\%~,
\label{CLEO}
\ee
\be
BR( B^0 \rightarrow J/\psi \bar K^0 )= (0.075 \pm 0.021) \% ~.
\label{CLEO0}
\ee
This failure, together with the scale problem pointed out above,
indicates that
factorization of matrix elements has to be accompanied by a
reinterpretation  of the Wilson coefficients. For class II
decays such as $B \ra J/\psi K$,
the short-distance $\mu$-dependent coefficient
$c_2(\mu) + c_1(\mu) / 3 $
has to be substituted by an effective coefficient $a_2$.
Phenomenologically \cite{BSW},
the latter is treated as
a free parameter to be determined from experiment. From
(\ref{factO}), (\ref{factoriz}),
and (\ref{CLEO}),
the most precise of the two measurements, one finds
\be
|a_2|= 0.32 \pm 0.02
\label{a2cleo}
\ee
where the quoted error is purely experimental.
The sign of $a_2$, of course, remains undetermined.

More theoretically, one can
argue that the above findings are a clear indication
for the existence of significant nonfactorizable contributions
which account for the difference of $a_2$ from the value
of $ c_2 +c_1/3$, by roughly
a factor of two at $\mu = m_b$, and which soften the strong
$\mu$-dependence.
A deeper analysis shows that the dominant effects should arise from
the nonfactorizable matrix element
of the operator  $\tilde{O}_2 $. Adopting
the parameterization
\be
\langle J/\psi K\mid \tilde{O}_2(\mu) \mid B\rangle =
2\tilde{f}(\mu)f_\psi m_\psi(\epsilon^\psi \cdot q) ~,
\label{nf}
\ee
one derives from (\ref{H}) and  (\ref{factoriz})
the effective coefficient
\be
a_2=c_2(\mu)+\frac{c_1(\mu)}3 +
2c_1(\mu)\frac{\tilde{f}(\mu)}{f_{B\ra K}^+}~.
\label{a2}
\ee
We consider the proof of this conjecture and the
direct calculation of $a_2$ (and of analogous
coefficients relevant for the other
classes of two-body decays \cite{BSW}) as one
of the most important tasks in heavy flavour physics.
As a first step in this direction,
we have undertaken a rough
estimate of $\tilde{f}$ using again QCD sum rules \cite{KLR}.
Following the general idea put forward in \cite{BS},
we work with the four-point correlation function
\be
<0\mid T\{j_{\mu5}^K(x)j_\nu^\psi(y)\tilde{O_2}(z)j^B_5(0)\}\mid 0> ~,
\label{corr}
\ee
where $ j_{\mu5}^K= \bar{u}\gamma_\mu \gamma_5s $ ,
$ j_\nu^\psi= \bar{c}\gamma_\nu c $, and
$ j^B_5= \bar{b}i\gamma_5 u $
are the generating currents of the mesons involved.

On the one side, one considers a dispersion
relation for this function in terms of intermediate hadronic states
in all three independent kinematical variables
(the squared momenta of the currents). Similarly as
in (\ref{disp}),
the ground state contribution
contains the matrix element of interest, that is
$\langle J/\psi K\mid \tilde{O}_2(\mu) \mid B\rangle $.
In addition, one has contributions from all kind of
excited resonances and continuum states with a very
complicated singularity structure.

On the other side, in the deep-euclidean region one can
apply a short-distance OPE to the correlation function (\ref{corr}).
This leads to a representation by a series of
vacuum expectation values (VEV) of local operators with calculable
coefficients. The VEV are universal nonperturbative
parameters characterizing the QCD vacuum. Best known examples
are the gluon and quark condensates.
We have performed this calculation including all
operators up to dimension 6.
More details of our analysis
are discussed in \cite{KLR}.
Here we only mention two complications that are
not present in the more familiar two- and three-point sum rules.
One problem  is the presence of a light continuum in the
$B$ channel below the $B$-meson pole.
This contribution can be associated with processes of the type
$`D^*D_s\mbox{{\it'}} \ra J/\psi K$
where an initial state carrying
$D^*D_s$ quantum numbers which is created by the combined
action of the operator product $\tilde{O}_2 j_5^B$ from
the vacuum  (in (\ref{corr}) ), rescatters into the final
$ J/\psi K$ state by strong interaction.
As a reasonable solution we suggest to
cancel this unwanted piece against a
corresponding term in the OPE of (\ref{corr}) which
develops a nonzero
imaginary part at $(p+q)^2 \geq 4m_c^2$  in the channel with the
corresponding quark content $c\bar{c}s\bar{q}$.
The second problem concerns the
subtraction of the contributions from higher resonances
and continuum states in the remaining sum rule.
Since the normal procedure (mentioned in section 2 )
does not work here, we use a
rough model for the spectral function including
besides the $B$, $J/\psi$ and $K$ ground states,  only the first
excited resonances.
We then perform a
Borel transformation in the $B$-meson channel and take
moments in the charmonium channel.
In the $K$-meson channel, we keep the momentum variable $q^2$
spacelike. We then fit the spectral function to
the sum rule
varying the Borel mass $M$, the moments and $q^2$. From this
fit we find
$\tilde{f} = -(0.045 \div 0.075)$, where the implicit scale $\mu$ is
identified with the central value
$M \simeq \sqrt{m_B^2-m_b^2}\simeq 2.5$ GeV.
Substituting this estimate in (\ref{a2}), and evaluating the
short-distance coefficients $c_{1,2}(\mu)$ also at $\mu= M$ instead
of the higher scale $\mu=m_b$ used in (\ref{BRfact}), we obtain
\be
a_2= -0.29 +0.38 -(0.19 \div 0.31) =-(0.10\div 0.22) ~,
\label{a2number}
\ee
where the three terms refer to the corresponding terms
in (\ref{a2}).

Several comments are in order. Firstly, the nonfactorizable matrix
element (\ref{nf})
is considerably smaller than the factorizable one
given in (\ref{factoriz}),
numerically, $|\tilde{f}/f^+_{B\ra K}(m_\psi^2)|\simeq 0.1 $.
Nevertheless, it has a strong quantitative impact on $a_2$
because of the large coefficient
$|2c_1/(c_2+c_1/3)| \simeq 20 \div 30 $.
Secondly, the factorizable
term $c_1/3$, nonleading in $1/N_c$,
and the nonfactorizable term proportional to $\tilde{f}$
are opposite in sign
and hence tend to cancel.
In fact, if
$|\tilde{f}|$ is taken at the upper end of the predicted
range, the cancellation is almost complete and
as a result increases the branching ratio
considerably. This is
exactly the scenario anticipated by the $1/N_c$-rule \cite{BGR}
and supported by similar estimates of nonfactorizable
amplitudes for $D$-decays \cite{BS}
and for $B\ra D\pi$ \cite{BS93,Halp}.
Thirdly, the theoretical value (\ref{a2number})
of $|a_2|$ is still somewhat
low as compared to
the phenomenological value (\ref{a2cleo}).
One should however note that the improvement due to the nonfactorizable
contribution
is remarkable as the
factorizable terms alone would give
$|a_2| = 0.09 $ for $\mu=M$.
Finally, our estimate
yields a negative overall sign for $a_2$ in contradiction to
strict factorization
(first two terms in (\ref{a2}))
and also to a global fit
of the data \cite{CLEO,HF}. We stress, however, that in this fit
the positive sign of $a_2$ is essentially determined from
the channels $B^- \ra D^0\pi^-, D^0 \rho^-$, $D^{*0} \pi^-$,
and $D^{*0}\rho^-$
and is then assumed to hold universally
for all channels. This assumption may not be
correct. Certainly,
the theoretical approach described here
provides no justification to expect universal values and/or
universal signs for the coefficients $a_{i}$ in
different channels. From
the relation (\ref{a2}) we see that universality can
at most be expected within certain classes of decay modes,
separately.
This point of view is also supported by the
phenomenological analysis of
nonfactorizable contributions in \cite{Soares}.

As a last remark we emphasize that
there is no simple relation between $B$ and $D$ decays
in our approach since the  OPE for the corresponding correlation
functions significantly differ in the relevant diagrams and in the
hierarchy of mass scales.
We hope to clarify these issues further.

\section{Conclusion}

QCD sum rule techniques are very useful in
calculating hadronic matrix elements for
exclusive heavy meson decays.
In particular, in combination with light-cone wave functions
for pions and kaons they provide a powerful tool to derive
heavy-to-light form factors and strong couplings of heavy to light
mesons. In contrast to other methods including HQET,
one can obtain
predictions for the whole range in momentum transfer, and also
for $D$ mesons which are presumably too light for
HQET to work. Moreover, finite mass effects are
automatically included.

In this paper, we have focussed on applications to the
CKM-suppressed semileptonic decays $ B \ra \pi l \nu_l$
and $D \ra \pi l \nu_l$, and to the factorization
approximation of the nonleptonic decay $B \ra J/\psi K$.

Using the latter mode as a prototype example,
we have discussed the shortcomings and principal problems
of factorization of nonleptonic
amplitudes. Again with the help of QCD sum rule
methods, we have directly estimated
the nonfactorizable contributions to the amplitude for
$B\ra J/\psi K $. Our results indicate that nonfactorizable effects
indeed play an essential role, putting the
$1/N_c$-rule to work similarly as in $D$-decays.
Moreover, on theoretical grounds the nonfactorizable
amplitudes are expected to be channel-dependent.
Hence, the pattern of two-body weak decays may be much richer than
what is revealed by the current phenomenological
analysis of the data.

\vspace{0.3cm}

{\bf Acknowledgements}.
R.R. thanks Neville Harnew and Peter Schlein for
the opportunity to participate in
this fruitful and enjoyable workshop.

\end{document}